\documentclass[seceq]{ptptex}

\usepackage{wrapft}
\usepackage{graphicx}

\def\cc{\c{c}}

\nofigureboxrule

\title{
PNJL model with eight quark interactions 
}

\author{%
A. H.~\textsc{Blin}$^{1}$,  
J.~\textsc{Moreira}$^{1}$,
A. A.~\textsc{Osipov}$^{1}$, B. ~\textsc{Hiller}$^{1}$ 
}

\inst{
$^1$Departamento de Fisica, Centro de Fisica Computacional, Faculdade de Ci\^encias e Tecnologia, Universidade de Coimbra, P-3004-516 Coimbra, Portugal\\
}

\abst{The phase diagram at finite temperature and chemical potential for the Polyakov-NJL model with eight quark interactions is reviewed for the $SU(3)$ flavor case. }

\begin{document}
\maketitle

A present day challenge to relativistic heavy ion experimental facilities is to establish whether a critical end point (CEP) can be identified for the hadron and quark gluon plasma phases of QCD, as predicted by chiral effective models and lattice QCD (LQCD).
At small baryonic chemical potentials $\mu_B$ LQCD seems to indicate that the transition occurs as a crossover implying that different observables may evidence different transition temperatures (denoted henceforth by $T_t$). The conjectured phase diagram combining current LQCD predictions and chemical freeze-out data \cite{Stephanov:2004,Gupta:2011} indicates that the CEP might occur at  $\frac{\mu_B}{T_c}\sim 2$ and $\frac{T}{T_c}\sim 1$, where $\mu_B=3 \mu$ and $\mu$ the quark chemical potential (taken for simplicity to be the same for all flavors).   
The present contribution aims at exploring the domains at which a CEP can occur within the NJL and PNJL models and in assessing $T_t$ for different observables at $\mu=0$. 
For the $SU(3)$ flavor NJL model combined with the $U_A(1)$ breaking 't Hooft interaction Lagrangian \cite{Nambu:1961}\tocite{Reinhardt:1988} the CEP has been calculated to be at $(\mu,T)=(324,48)$MeV and for the PNJL model at $(\mu,T)=(313,102)$MeV  \cite{Fukushima:2008}, yielding $\frac{\mu_B}{T_c}\sim 20$ and $\sim 9$ respectively.  
Other variants of the NJL model, such as the $SU(2)$ flavor case \cite{Kashiwa:2008}, or with inclusion of the vector mesons \cite{Asakawa} all have the fact in common that they lead to a CEP at $\mu > 300$MeV. 
This picture changes drastically if one includes the $8q$ interaction vertices \cite{Osipov:2006b}, which allow for  $\frac{\mu_B}{T_c}\sim 2$, see Fig.1. One combination violates the OZI rule and has remarkable implications. Its strength $g_1$ can be increased combined with a decrease of the $4q$ coupling magnitude $G$, almost without changing the  low meson spectra in vacuum  \cite{Osipov:2007a,Hiller:2010}. At finite T an increase of $g_1$ leads to the following changes in the phase diagram of the model: (i) at $\mu=0$ the $T_t$ at which chiral symmetry is restored is substantially lowered \cite{Osipov:2007b,Osipov:2008}, in accordance with LQCD \cite{Aoki:2006}. (ii) From a certain value of $g_1$ the location of the CEP is associated with a pattern of dynamical symmetry breaking induced by the 't Hooft interaction strength rather than by the G coupling \cite{Osipov:2008,Hiller:2010}. (iii) The CEP is shifted to higher $T$ and lower $\mu$ \cite{Hiller:2010}. (iv) The sharpness of crossover transitions is associated with the strength $g_1$. (v) 8q interactions lead to a suppression of the fictitious quark degrees of freedom of the NJL model in the crossover region \cite{Hiller:2010}, adding to the suppression observed for the PNJL model \cite{Fukushima:2008}.
Since we have introduced the $8q$ interactions for the $SU(3)$ flavor case \cite{Osipov:2006b}, their implications have been intensely studied by other groups as well, to analyze several aspects of the phase diagram of the $SU(2)$ flavor  PNJL and dressed Polyakov loop model \cite{Kashiwa:2008,Kashiwa:2008a}, also in a strong magnetic background field \cite{Gatto}.  The $SU(3)$ flavor PNJL model has been considered in \cite{Bhattacharyya:2010}. 

In Fig.1 bold solid lines are for 1st order transitions for the weak  $8q$ coupling $g_1=1000$GeV$^{-8}$ (upper curve) and strong $g_1=8000$GeV$^{-8}$ (lower curve).  At $\mu=0$  the $T_t$ diminishes considerably with increasing $g_1$. The $\mu$ is pushed towards lower values with increasing $g_1$, at the CEP (circles) for strong $g_1$ it is reduced by $\sim$ half as compared to the weak case. 
Dotted lines stand for the crossover regime, dashed lines are spinodals. 
\vspace*{-0.4cm}

\begin{figure}[htb]
\parbox{\halftext}{
\centerline{\includegraphics[width=6.5 cm,height=4. cm]
{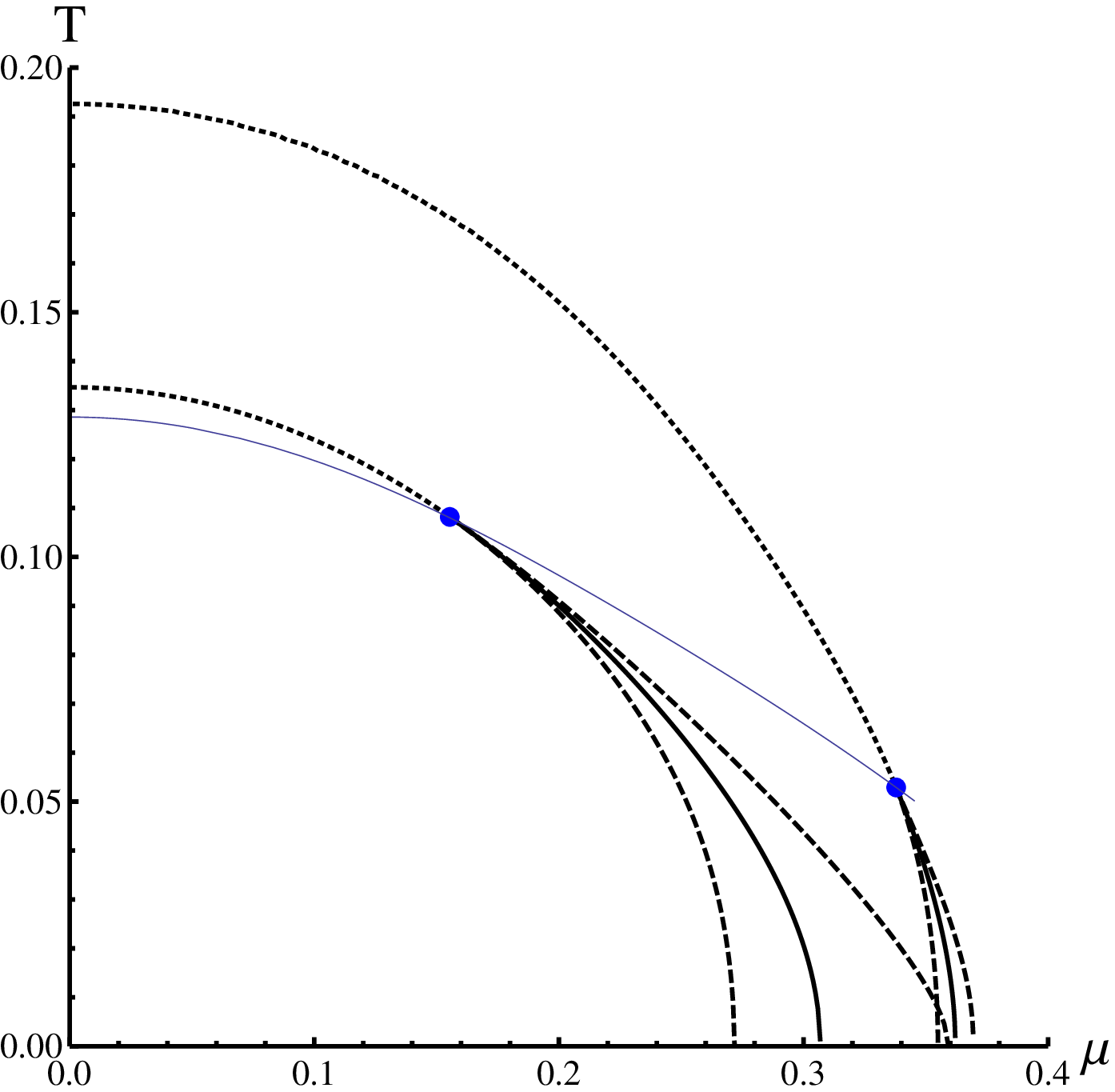}}
\figurebox{6cm}{0cm}
\caption{Phase diagram for the $SU(3)$ flavor NJL model\cite{Hiller:2010}, see text for description. 
The thin line connects CEP obtained by varying the strength of $g_1$.}}
\hfill
\parbox{\halftext}{
\centerline{\includegraphics[width=6.5 cm,height=4. cm]
{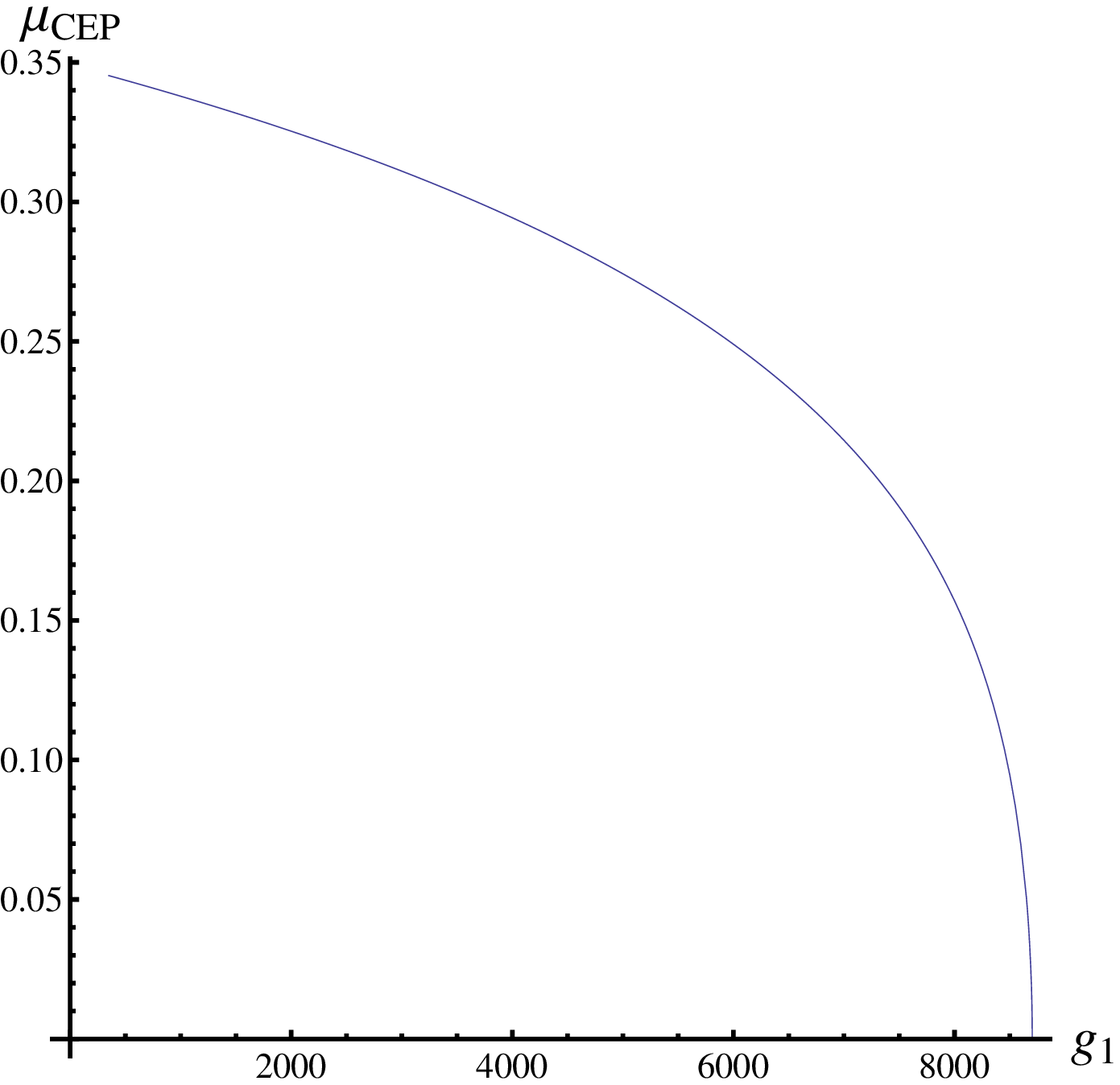}}
\figurebox{6cm}{0cm}
\caption{The quark chemical potentials $\mu_{CEP}$ as function of the $8q$ strength $g_1$, correponding to the thin line of CEP in Fig.1. 
\vspace*{0.2cm}
}}
\end{figure}
\vspace*{-0.6cm}

\begin{figure}[htb]
\parbox{\halftext}{
\centerline{\includegraphics[width=6.5 cm,height=4. cm]
{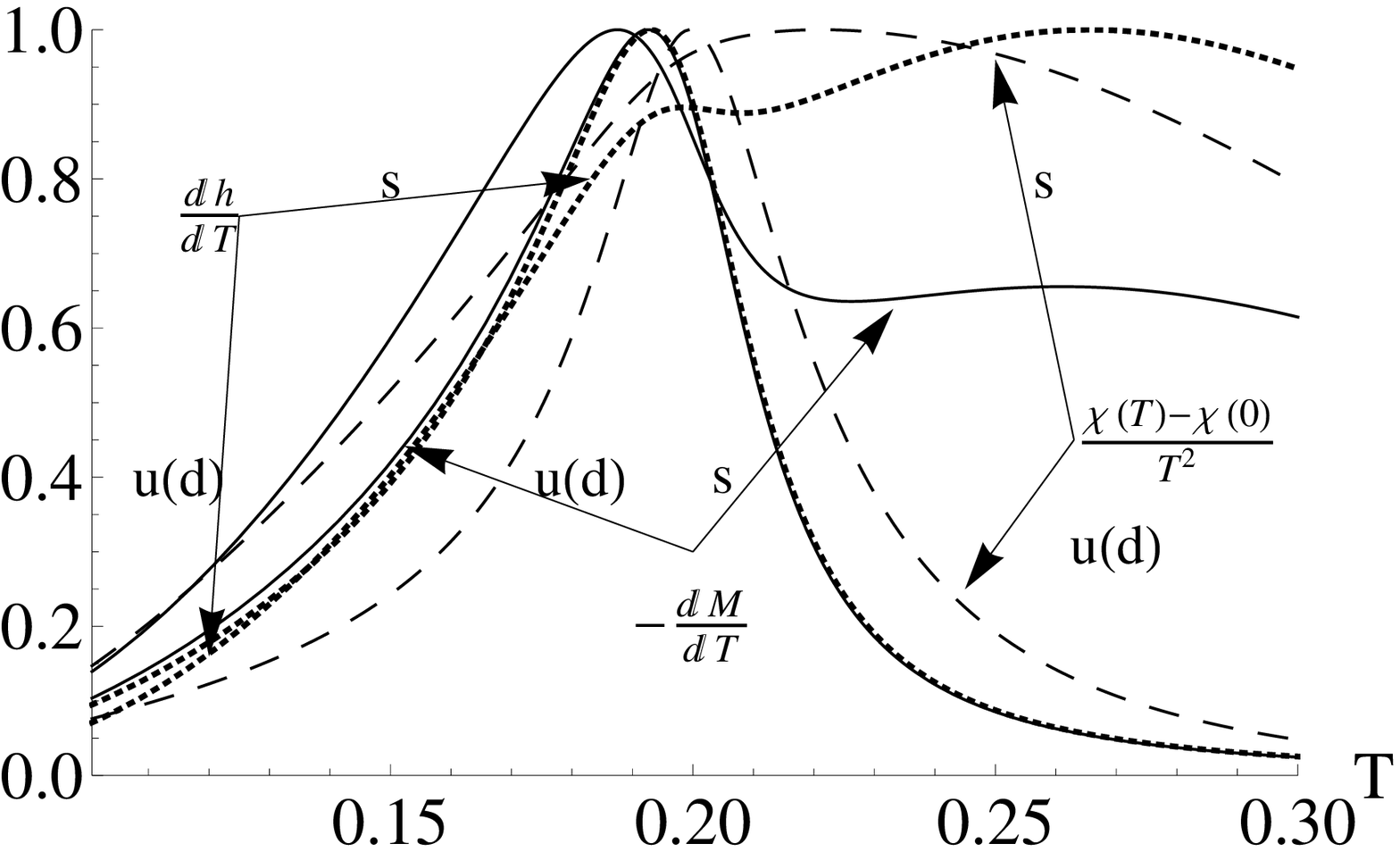}}
\figurebox{6cm}{0cm}
\caption{The chiral transition crossover temperatures $T_t$ in the $SU(3)$ flavor NJL, given by the peaks of the observables, in the weak $8q$ regime (see text).}}
\hfill.
\parbox{\halftext}{
\centerline{\includegraphics[width=6.5 cm,height=4. cm]
{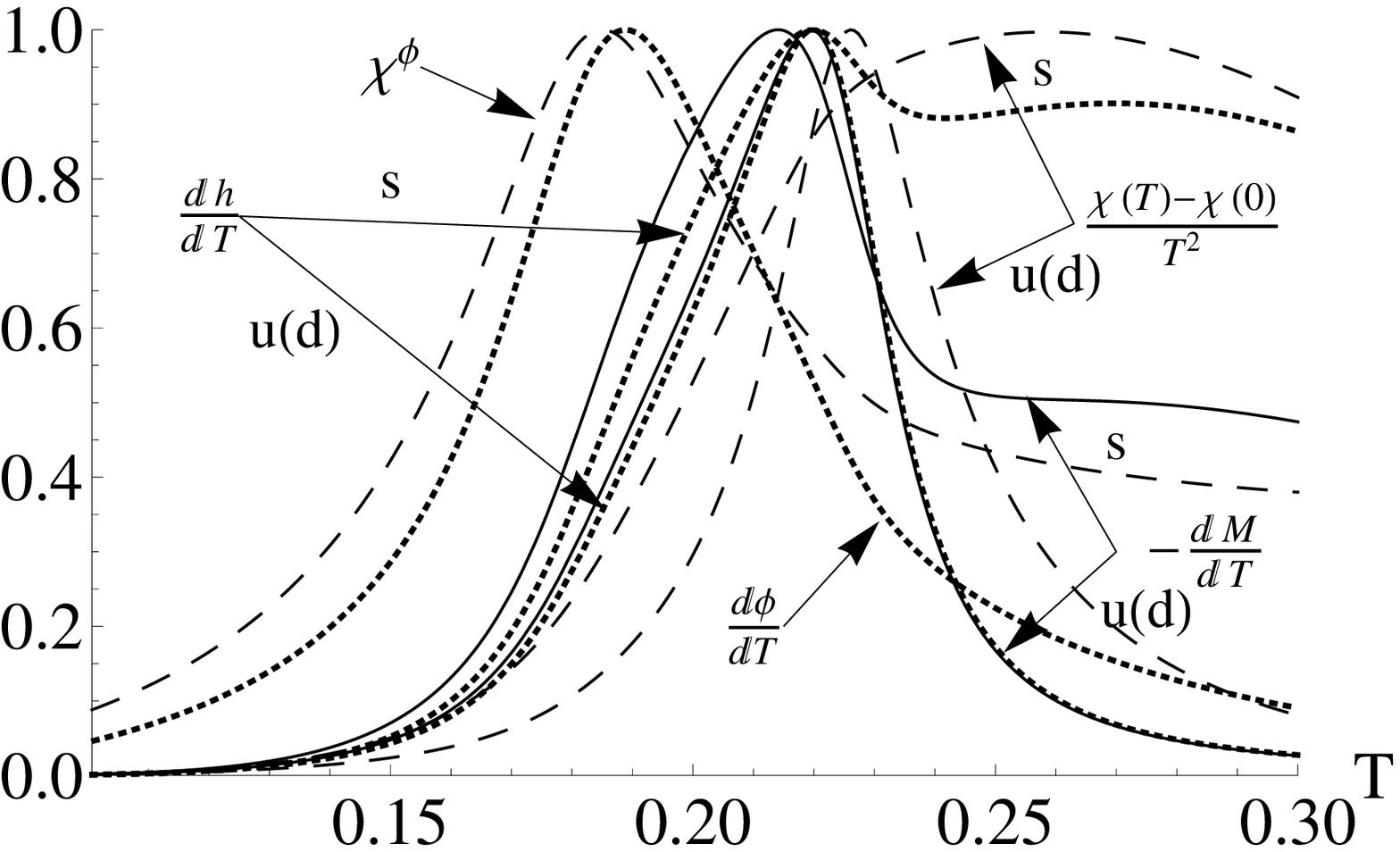}}
\figurebox{6cm}{0cm}\caption{The same as in Fig.3 in the case of the PNJL model. The deconfinement $\chi^{\Phi}$ peak occurs before the chiral crossovers. \vspace*{0.2cm}}}
\end{figure}

Figs. 3 and 5 show $T_t$ at $\mu=0$  (at maximum of the curves), for observables of the $SU(3)$ NJL model at weak  $g_1=1000$GeV$^{-8}$ and strong $g_1=6500$GeV$^{-8}$ couplings (parameters from sets (a), (b) of \cite{Moreira:2011}, with $G=7.865$GeV$^{-2}$ for the strong case). The curves are normalized to unity at the peaks. Figs. 4 and 6 show the same for the PNJL model (Polyakov potential from \cite{Ratti:2006} with $T_0=210$MeV). For the weak coupling one sees that different definitions lead to a spread in $T_t$, especially for the strange quark: the inflection points for the quark masses and condensates, given by the maxima of their T derivatives, differ considerably from the normalized strange quark susceptibility $\frac{\chi(T)-\chi(0)}{T^2}$ with $\chi(T)=-\frac{\partial^2 \Omega(T)}{\partial m_s^2}$ \cite{Fukushima:2008,Min:2009,Bazavov:2010} and $m_s$ denoting the current strange quark mass. The peak of the Polyakov loop susceptibility \cite{Fukushima:2008} $\chi_{\Phi}=T^2 \frac{\partial^2 (-\Omega)}{\partial \eta \partial \eta'}$   is a measure for the deconfinement $T_t$, which is smaller than the chiral $T_t$ for all cases considered. The thermodynamic potential $\Omega$ is derived in \cite{Hiller:2010} and \cite{Moreira:2011}. 
\vspace*{-0.6cm}

\begin{figure}
\parbox{\halftext}{
\centerline{\includegraphics[width=6.5 cm,height=4. cm]
{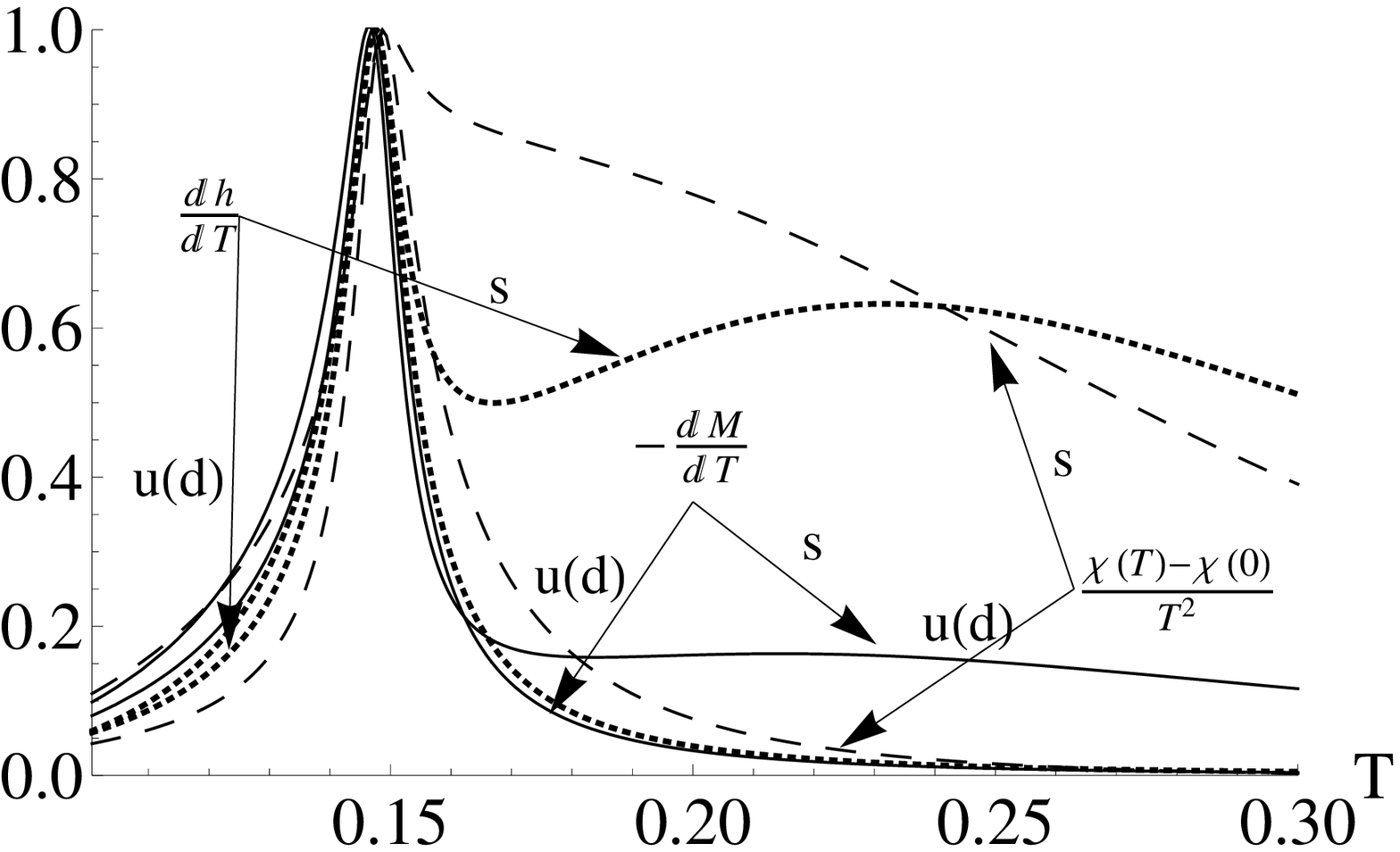}}
\figurebox{6cm}{0cm}
\caption{The same as in Fig.3 in the strong $8q$ regime, 
(see text). The chiral crossover converges for all observables considered. }}
\hfill
\parbox{\halftext}{
\centerline{\includegraphics[width=6.5 cm,height=4.5 cm]
{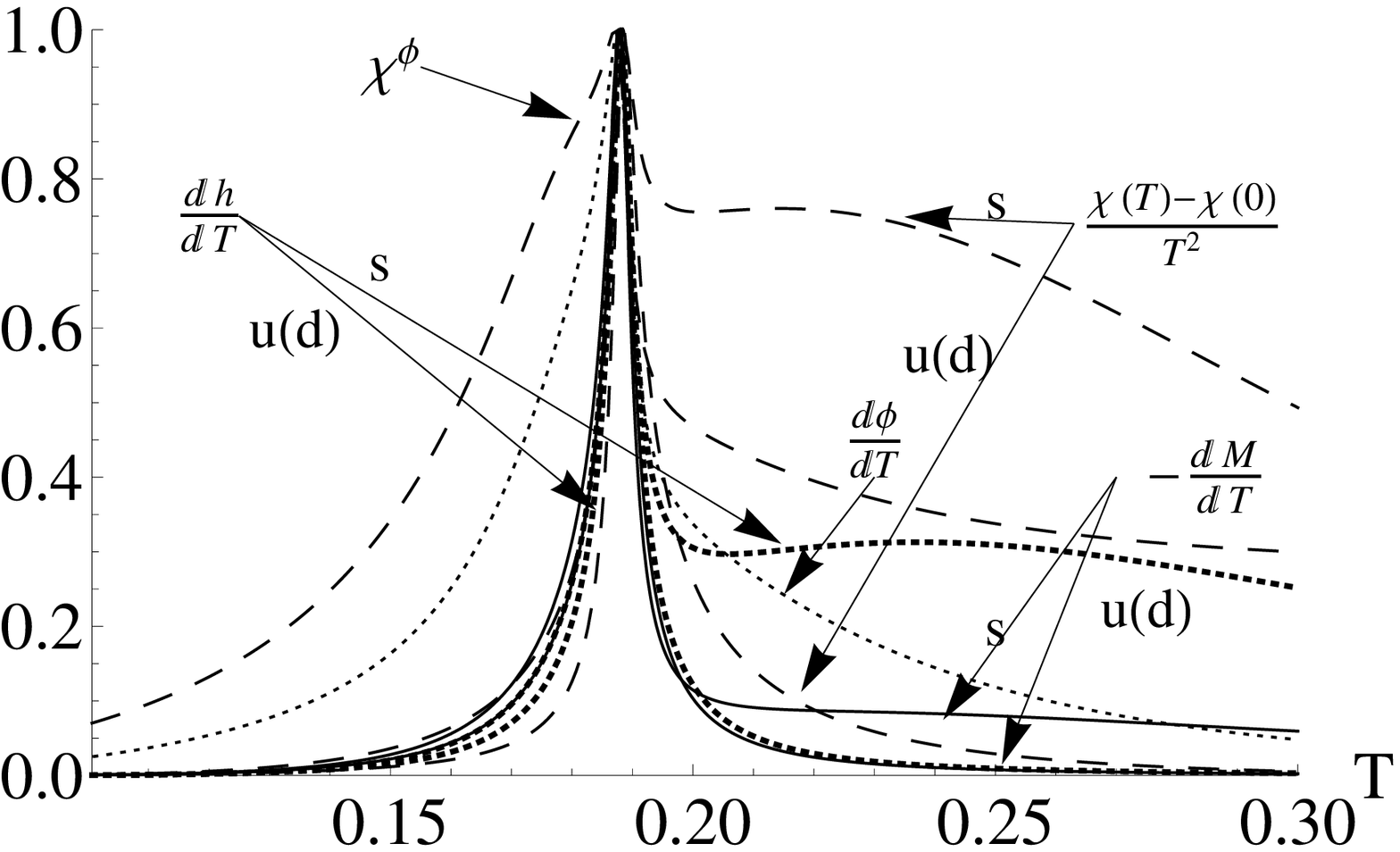}}
\figurebox{6cm}{0cm}\caption{The same as in Fig.5 in the case of the PNJL model. The deconfinement $\chi^{\Phi}$ peak converges to the chiral crossover peaks.\vspace*{0.5cm}
}}
\end{figure}
\vspace*{-0.5cm}


{\small {\bf Acknowledgements} We acknowledge with great pleasure  the interest of Profs. A. Nakamura, A. Ohnishi an W. Florkowski in our work. Supported by Funda\cc \~ao para a Ci\^encia e Tecnologia,
SFRH/BPD/63070/2009, CERN/FP/116334/2010 and Centro de F\'isica Computacional, unit 405.}

\end{document}